\documentclass[apj,revtex4]{emulateapj}

\newcommand{\kms}{km$~$s$^{-1}$}
\newcommand{\vsini}{$v \sin i$}

\newcommand{\ali}{$A$(Li)}

\newcommand{\cratio}{$^{12}$C/$^{13}$C}
\newcommand{\msun}{$M_{\sun}$}

\newcommand{\teff}{$T_{\rm eff}$}

\shortauthors{Carlberg et~al.}

\begin{document}

\title{Lithium in Open Cluster Red Giants Hosting Substellar Companions}

\author{Joleen K.\ Carlberg\altaffilmark{1,4}, Verne V.\ Smith\altaffilmark{2}, Katia Cunha\altaffilmark{3}, Kenneth G.\ Carpenter\altaffilmark{1}}
\altaffiltext{1}{NASA Goddard Space Flight Center,
Code 667,
Greenbelt, MD 20771, USA
}
\altaffiltext{2}{National Optical Astronomy Observatory, 950 North Cherry Avenue, Tucson, AZ 85719, USA}
\altaffiltext{3}{Observat\'orio Nacional, Rua General Jos\'e Cristino, 77, 20921-400 S\~ao Crist\'ov\~ao, Rio de Janeiro, RJ, Brazil}
\altaffiltext{4}{NASA Postdoctoral Program Fellow,  joleen.k.carlberg@nasa.gov}

\begin{abstract}
We have measured stellar parameters, [Fe/H], lithium abundances, rotation,  and \cratio\ in a small sample of red giants in three open clusters that are each home to a  red giant star that hosts a substellar companion (NGC2423~3, NGC4349~127, and BD+12~1917 in M67).  Our goal is to explore whether the presence of substellar companions influences the Li content. Both \cratio\ and stellar rotation are measured as additional tracers of stellar mixing.  One of the companion hosts,  NGC2423~3,  is found to be Li-rich with \ali$_{\rm NLTE}=$1.56~dex, and this abundance is significantly higher  than the \ali\ of the two comparison stars in NGC~2423.    All three substellar companion hosts  have the highest \ali\  and \cratio\ when compared to the control red giants in their respective clusters; however, except for NGC2423~3, at least one control star has similarly high abundances within the uncertainties. Higher \ali\ could suggest that the formation or presence of planets  plays a role in the degree of internal mixing on or before the red giant branch.    However, a multitude of factors affect \ali\ during the red giant phase, and when the abundances of our sample are compared to abundances of red giants in other open clusters available in the literature, we find that they all fall well within a much larger distribution of   \ali\ and \cratio. Thus, even the high Li in NGC2423~3 cannot be concretely tied to the presence of the substellar companion.  
\end{abstract}


\section{Introduction}
The abundance of lithium, \ali, in stellar photospheres is a useful diagnostic of a variety of phenomena related to the internal mixing near the stellar surface.  It has been used as a crude clock for some low mass main sequence (MS) stars (see \citealt{sesito05} for uses and limitations) 
and to study non-standard, i.e., non-convective, mixing processes in red giant (RG) stars (\citealt{Gratton:2000wg}, \citealt{2003ApJ...593..509D}, \citealt{2009PASA...26..168G}). Moreover,  studies of MS stars  suggest that the planet formation process may alter the stellar \ali\ early in the star's life by causing enhanced Li-depletion (e.g., \citealt{Israelian:2009fj}, \citealt{2014MNRAS.441.1201G}, and references therein), which is supported theoretically \citep[e.g.,][]{2010A&A...521A..44B}. However, whether \ali\  of MS  planet hosts truly differs from apparently planet-free stars after controlling for stellar ages, masses, etc.\ is still hotly debated (e.g, \citealt{2010ApJ...724..154G}, \citealt{2010A&A...519A..87B},  \citealt{2012ApJ...756...46R}). %  

Because the number of known RGs that host planets or other substellar companions (SSCs) lags considerably behind the number of MS hosts, there is not much known about whether Li in planet hosts is substantially different from non-planet hosts. Ten  RG planet hosts were included in the \cite{carlberg12} study of Li and rotation in RGs, and the planet hosts did not show any unusual \ali.   Two SSC hosts were included in a contemporaneous study of \ali\ in open clusters, and these appear to have enhanced \ali\ \citep{mena15}.
Most recent studies comparing the abundance patterns of RG planet hosts and non-hosts have not included Li in their analyses (e.g., \citealt{2013A&A...554A..84M} and \citealt{2015A&A...574A..50J}). 

Complicating any comparisons between  SSC hosts and control populations are  the difficulty in estimating accurate masses for RGs and the large dispersion of \ali\ seen in RGs---at least three orders of magnitude for ``normal'' RGs and two more orders of magnitude when including the  rare Li-rich stars.  Li-rich stars, generally considered to be those with \ali$\gtrsim1.5$~dex, exceed the Li  abundances  predicted by standard models,  and the addition of non-standard mixing tends to predict lower \ali\ (e.g., \citealt{1995ApJ...453L..41C}, \citealt{2003ApJ...593..509D}) except in special cases (\citealt{1999ApJ...510..217S}, \citealt{2001A&A...375L...9P}, \citealt{2004ApJ...612.1081D}).
 Other physical properties of red giants that are associated with  Li-richness include fast rotation (e.g., \citealt{2002AJ....123.2703D} and \citealt{carlberg12}) and
 infrared excesses (e.g., \citealt{2015ApJ...806...86D} and references therein).
\cite{1999MNRAS.308.1133S} suggested that high Li, fast rotation, and infrared excess could all  be plausibly explained by the engulfment of planets or brown dwarfs.
At least one planet-hosting RG in the field is known to be Li-rich \citep{Adamow:2012ii}.  

 Open clusters have historically been used to understand the mixing processes that affect the abundances of Li and other elements  during the red giant phase, especially to discern the mass dependence. 
  Mixing in RGs not only dilutes Li but also alters the surface C, N, and O abundances and reduces \cratio. 
While the first evidence of non-standard mixing in Population~I stars was found for field RGs  (e.g., \citealt{1976ApJ...210..694T} and references therein),
\cite{gilroy89} was the first to study the mass dependence using open clusters. 
She found a positive correlation between \cratio\ and the cluster turn-off mass up to 2.2~\msun\ but no correlation between \ali\ and the turn-off mass.   
\cite{2009A&A...502..267S} found the same correlation between \cratio\ and turn-off mass and also identified a small grouping of massive stars with low \cratio. They interpreted this as a signature of extra mixing taking place during the early-AGB stage.   

Extra mixing also occurs in low mass stars at the luminosity bump phase (the stage when the H-burning shell encroaches on the region of the star that was well mixed by convection during first dredge-up) where additional Li depletion occurs via thermohaline mixing \citep{2007A&A...467L..15C}. 
Evidence of this mixing is seen in the dichotomy of \ali\ in open clusters in the region of the color-magnitude diagram (CMD) where red clump stars and first ascent stars overlap (e.g., \citealt{pilachowski88}, \citealt{2001A&A...374.1017P}, \citealt{2004AJ....127.1000A}, \citealt{2004A&A...424..951P}, \citealt{2009AJ....138.1171A} )

  Open clusters were also instrumental in progressing our understanding of a mass-dependent Li-depletion mechanism  in MS stars, known as the Li-dip. This feature was first found in the Hyades \citep{1986ApJ...302L..49B}, where MS stars in a very narrow temperature range show a large depletion in Li compared to 
 stars that are both hotter and cooler. 
 \cite{2012AJ....144..137C}  showed  that the mass at which the Li-dip occurs increases with increasing metallicity.
 Understanding the amount of Li depletion on the main sequence is of course necessary for interpreting the abundances measured at later stages of evolution.

 The goal of this study is to test whether the presence of substellar companions affects the level of \ali\ observed during the red giant phase. 
Given the complex evolution of Li throughout a stars's life, we focused the study
 on three substellar companion-hosting RGs (2 planetary mass and 1 brown dwarf mass companion) that are members of open clusters.
Cluster membership allows for a more precise estimate of the stellar masses and, importantly, the comparison to other RGs of nearly identical mass and composition,  which provides a better opportunity for isolating effects due to the presence of companions.  Therefore, we 
 also present \ali\ for 1--3 red giants in each cluster that are not known to host planets.  
We also measure \cratio\  and \vsini\ to explore possible differences in the  stars' mixing histories.

\section{Clusters Properties and Stellar Sample}
\label{sec:properties}
Most of the stars in this study come from  \cite{2007A&A...472..657L}, which describes their program for monitoring RGs in 13 open clusters for planetary companions. Their sample includes 6 stars in NGC~2423 (3, 43, 240, 73, 20, and 56) and 7 stars in NGC~4349 (127, 174, 53, 168, 203, 5, and 9).
In both clusters, a SSC was discovered around the brightest (presumably most evolved) star monitored in each cluster.
Two additional stars come from \cite{2014A&A...561L...9B}, who discovered three planet-hosting stars (two MS stars and one RG)  among a large sample of M67 stars being monitored for companions. Again, the planet was discovered around one of the most evolved stars monitored.

\begin{figure}[tb]
\includegraphics[width=0.27\textwidth,angle=90]{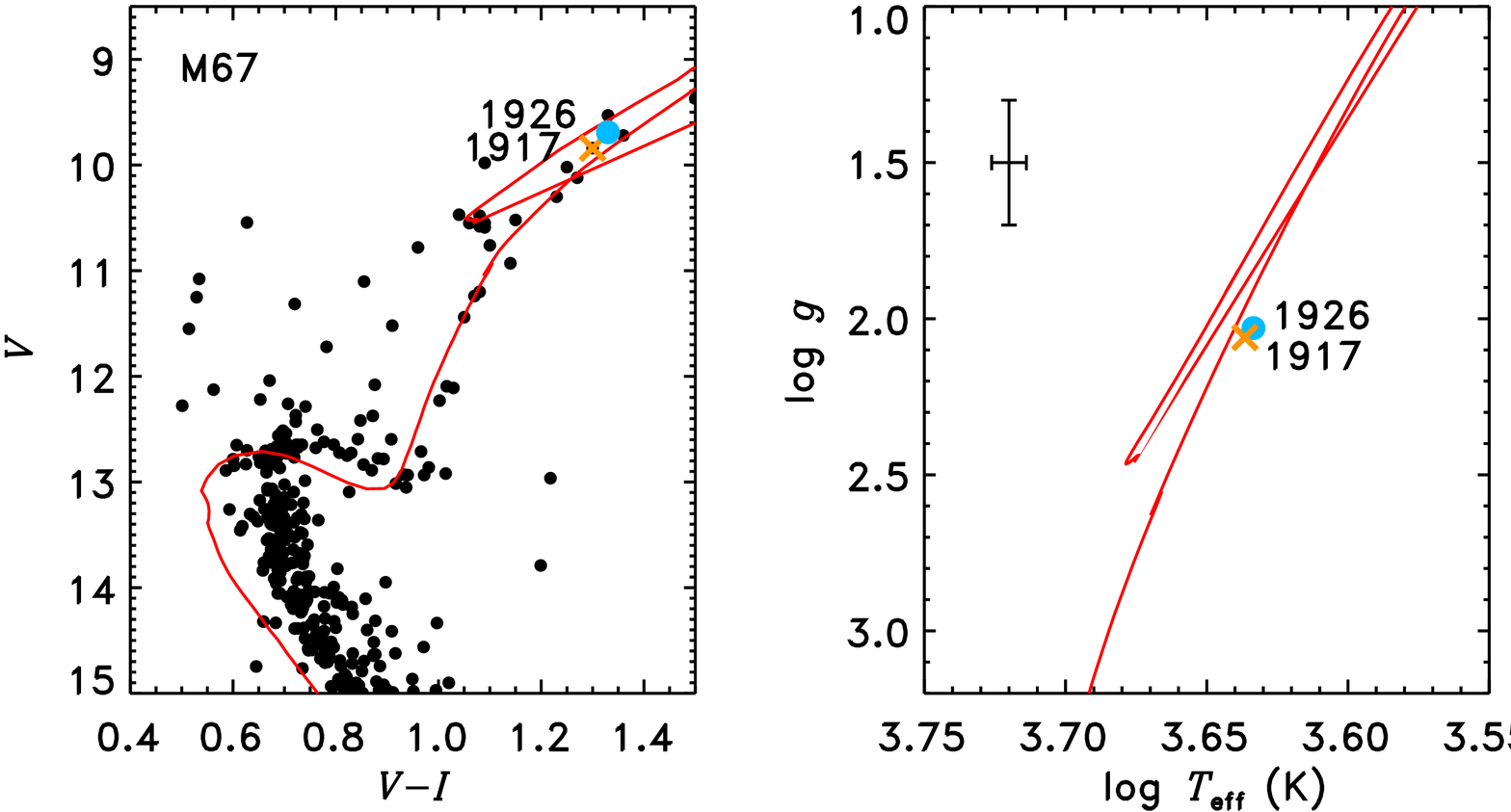} 
\includegraphics[width=0.27\textwidth,angle=90]{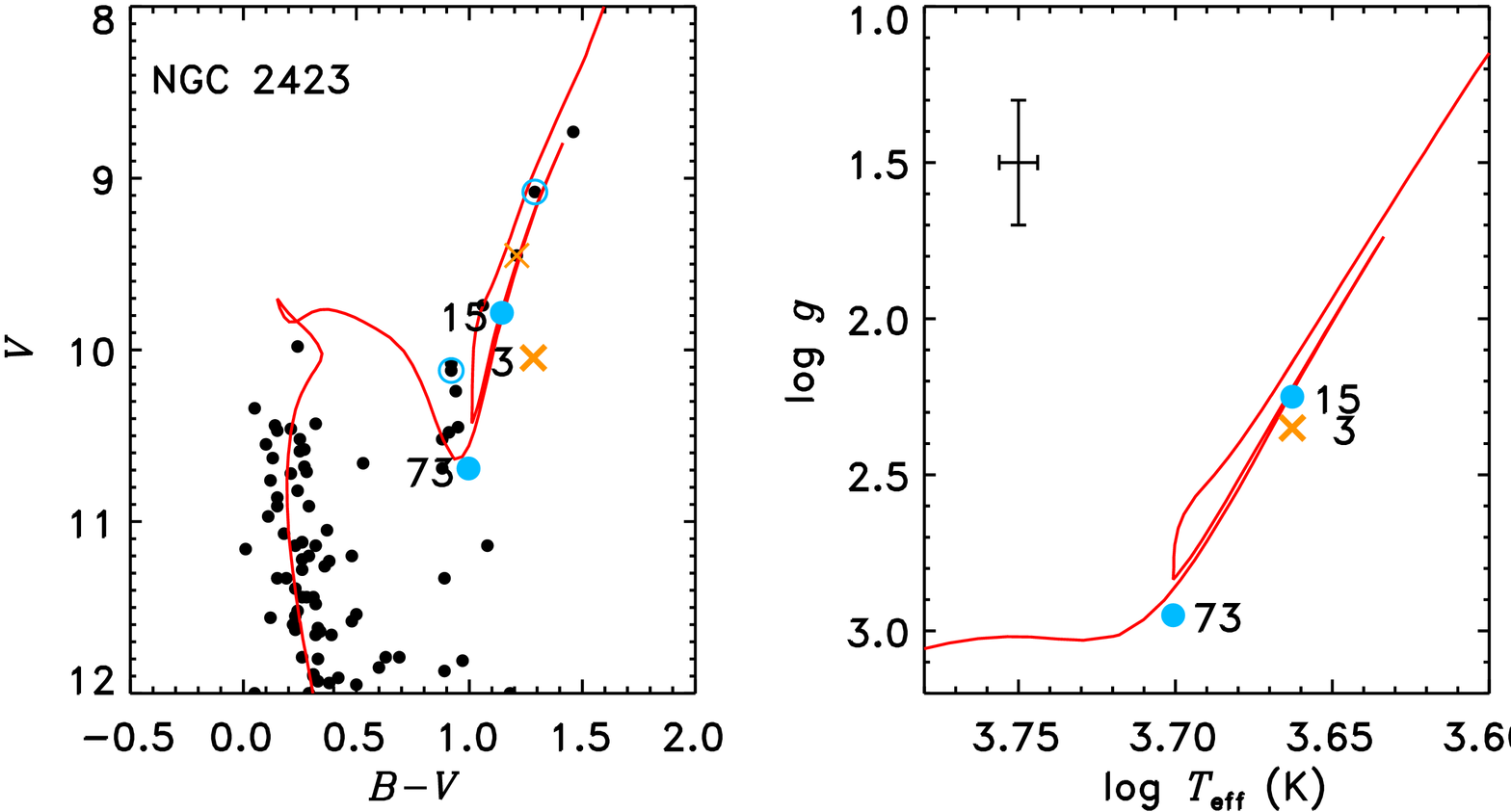} 
\includegraphics[width=0.27\textwidth,angle=90]{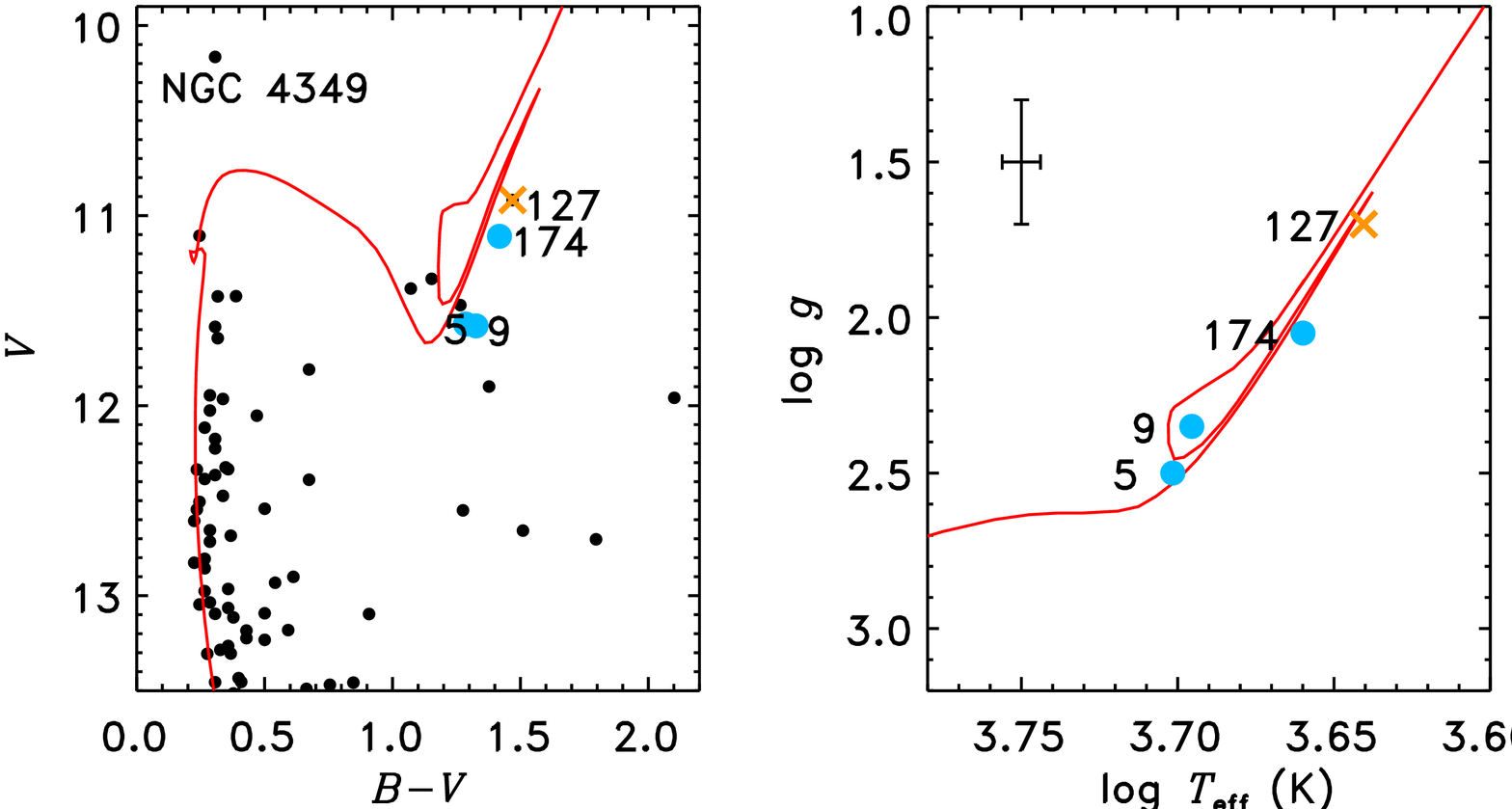} 
\caption{ Optical CMDs (left column) and \teff-$\log g$ diagrams (right column) of  M67 (top row), NGC~2423 (middle row), and NGC~4349 (bottom row).  The $\times$ is the SSC hosting star in each cluster, and the large, blue circles are the comparison stars.  
Photometry for M67 comes from \cite{1993AJ....106..181M}. 
The photometry for NGC~2423 comes from \cite{1976A&AS...26...13H}, with the  thin-lined/open symbols for the target stars.  The thicker/filled symbols of the target stars correspond to the photometry from \cite{2000A&A...355L..27H}.
Photometry for NGC~4349 comes from \cite{1961AN....286..105L}, after applying the corrections  suggested by \cite{2012A&A...537L...4M}.
\label{fig:cmds}}
\end{figure}

M67 is one of the most well-studied open clusters and has a near-solar age and metallicity.  
We adopt an age of 4.3~Gyr, $E(B-V)=0.04$,   and $d=832$~pc for M67 from  \cite{2004A&A...414..163S} and  [Fe/H]=$-0.06$ (see Section \ref{sec:stell_param}). 
A \cite{2012MNRAS.427..127B} isochrone with these properties gives the masses of the stars on the red giant branch (RGB) to be $\sim$1.3~\msun, consistent with the mass adopted in \cite{2014A&A...561L...9B} for the planet hosting star ($1.35\pm0.05$~\msun). In the top left column of Figure \ref{fig:cmds}, we show the CMD of M67 with photometry from \cite{1993AJ....106..181M} and the \cite{2012MNRAS.427..127B} isochrone with the adopted cluster parameters. The planet hosting star is BD+12~1917, and we selected BD+12~1926 as a comparison star for its very similar color and magnitude. 

The properties of the other two clusters are more disputed.   \cite{2007A&A...472..657L}  adopted the cluster parameters available in WEBDA for NGC~2423, and we also adopt those values for the age ($\log \tau$=8.867),  reddening (E$(B-V$)=0.097), and distance ($d$=766 pc).  They adopted [Fe/H] of $+0.14\pm0.09$ from \cite{1997AJ....114.2556T}.  However, more recent studies find lower metallicities (\citealt{2013A&A...557A..70M} and   \citealt{2014A&A...561A..93H}). Our measured [Fe/H] is $-0.01$~dex.   In the middle left panel of Figure \ref{fig:cmds}, we show optical photometry \citep{1976A&AS...26...13H} of the cluster together with an isochrone adopting the metallicity of $-0.01$~dex.
The planet host is NGC~2423~3, and the two comparison stars were selected to bracket that star in magnitude.  The thin-lined/open symbols 
of the three RGs we target show the photometry of \cite{1976A&AS...26...13H}, while the thicker/filled symbols show the updated photometry of these stars \citep{2000A&A...355L..27H} used in Section \ref{sec:masses}.

For NGC~4349,  the available WEBDA values are  $\log \tau$=8.315,  $E(B-V)$=0.384, and $d$=2176 pc.   \cite{2007A&A...472..657L} adopted  [Fe/H]=$-0.12\pm0.04$~dex from \cite{1995AJ....110.2813P}.  Newer studies make substantial changes to these parameters. 
 \cite{2012A&A...537L...4M} re-derived cluster properties using infrared photometry and found the cluster to be  older ($\log \tau$=8.55), closer ($d$=1630~pc) and less-reddened ($E(B-V)$=0.291). Their results were insensitive  to  adopted metallicity. We find a metallicity of [Fe/H]=$-0.15$~dex. The bottom left panel of Figure \ref{fig:cmds}  shows the optical photometry of \cite{1961AN....286..105L}, corrected to the standard photometry system with the corrections given by \cite{2012A&A...537L...4M}.  The isochrones adopt the \cite{2012A&A...537L...4M} properties and our metallicity.  
 The SSC host is the brightest RG, NGC~4349~127, and we selected three apparently less-evolved RGs as comparison stars.  \cite{2012A&A...537L...4M} revised the mass measurement of  NGC~4349~127 from  $3.9 \pm 0.3$~\msun\  \citep{2007A&A...472..657L} down to  $\sim 3.1$~\msun,  which would lower the companion's $m \sin i$ from 19.8 to 17.0~$M_{\rm Jup}$.   

\section{Observations}
All of the stars analyzed in this study were observed with the Magellan Inamori Kyocera Echelle (MIKE) spectrograph between May~2012 and March~2014 using the $0.5\arcsec$ slit, which results in a resolving power of $R_{\lambda}\sim$~44,000.
The data were reduced with the Carnegie python pipeline \citep{2003PASP..115..688K}\footnote{Available at \url{http://code.obs.carnegiescience.edu/mike}}, which completes standard CCD processing (overscan
subtraction, bad pixel masks, flat fielding, sky subtraction), and extracts and wavelength calibrates  the target spectra.  Only data from the red arm of the spectrograph, covering 4800--9400~\AA, were used in this analysis.  The 34 echelle orders were combined to a one-dimensional format and velocity shifted to the stellar rest frame.   The signal-to-noise ratio (S/N) of the data is 150--250 near 6700~\AA.

\section{Stellar Parameters and Abundances}
\subsection{\teff, $\log g$, [Fe/H], \& $\xi$}
\label{sec:stell_param}
The stellar parameters of \teff, $\log g$, [Fe/H], and $\xi$ were measured spectroscopically using  equivalent widths (EWs) of \ion{Fe}{1} and \ion{Fe}{2} lines. We used the \cite{carlberg12} Fe line list, and measured EWs for 72 \ion{Fe}{1} and 13 \ion{Fe}{2} lines.  These measurements  are given in Table \ref{tab:eqws}.
 We used the  `abfind' driver in the 2014 release of MOOG \citep{sneden73} to compute $A$(Fe) from the EWs, using MARCS spherical atmosphere models (\citealt{2008A&A...486..951G}, \citealt{2008PhST..133a4003P}).
 The parameters of the atmospheric models were adjusted until  the output 
$A$(\ion{Fe}{1}) and $A$(\ion{Fe}{2}) were within 0.02~dex and  there was no correlation between the output $A$(\ion{Fe}{1})  and either the lines' excitation potentials or reduced EWs.
The results are given in the first six columns of Table \ref{tab:results}.  The $\sigma_{\rm [Fe/H]}$ column is the standard deviation of the \ion{Fe}{1} lines.

We find average cluster metallicities of [Fe/H]~$=-0.01$~dex ($\sigma=0.02$) for NGC~2423, [Fe/H]~$=-0.15$~dex ($\sigma=0.06$) for NGC~4349, and [Fe/H]~$=-0.06$~dex ($\sigma=0.00$) for M67. Our measured $T_{\rm eff}$ and $\log g$ are compared to the cluster isochrones in the right panels of Figure \ref{fig:cmds}. A representative error bar is given in each panel. The stellar parameters are in excellent agreement with the cluster isochrones. 

 In Table \ref{tab:comp}, we compare our measurements of the stellar parameters of the  SSC hosts to the values reported in the literature.  
\cite{2010ApJ...725..721G} and \cite{2013A&A...557A..70M} measured parameters for two of the stars.
For NGC 2423 3, our result is intermediate to these two studies.  For NGC 4349 127, our \teff\ and $\xi$ are lower than both studies, but generally agree  within the uncertainties. 
For BD+12~1917, all of our parameters agree with \cite{2014A&A...561L...9B} within the uncertainties. 

\begin{deluxetable*}{crrrrrrrrrrrrr}
\tablecolumns{13}
\tablewidth{0pt}
\tabletypesize{\scriptsize}
\tablecaption{Iron Line Equivalent Widths \label{tab:eqws}}
\tablehead{
   \colhead{$\lambda$} &
   \colhead{Species} &
   \colhead{$\chi$}&
   \colhead{$\log gf$} &
   \colhead{EW$^1$} &
   \colhead{EW$^2$} &
   \colhead{EW$^3$} &
   \colhead{EW$^4$} &
   \colhead{EW$^5$} &
   \colhead{EW$^6$} &
   \colhead{EW$^7$} &
   \colhead{EW$^8$} &
   \colhead{EW$^9$} \\
   \colhead{(${\rm \AA}$)} &
   \colhead{  } &
   \colhead{ (eV)} &
   \colhead{(dex)} &
   \colhead{(m\AA) } &
   \colhead{(m\AA) } &
   \colhead{(m\AA) } &
   \colhead{(m\AA) } &
   \colhead{(m\AA) } &
   \colhead{(m\AA) } &
   \colhead{(m\AA) } &
   \colhead{(m\AA) } &
   \colhead{(m\AA) } &
 }
\startdata
5307.36 & 26.0 &  1.61  &  -2.99 & 185.2 & 186.1 &  172.1 &  176.7 & 147.6 & 154.9 & 150.4 & 189.6  & 179.4\\
5322.04 & 26.0 &  2.28  &  -2.80 & 134.65& 138.4 &  128.6 &  130.0 & 106.2 & 112.9 & 105.1 & 143.7  & 132.8\\
5466.99 & 26.0 &  3.57  &  -2.23 & 91.9  & 91.9  &  83.9  &  88.9  & 66.8  & 68.2  & 65.3  & 90.9   & 84.9\\
5491.83 & 26.0 &  4.19  &  -2.19 & 42.9  & 43.6  &  43.2  &  45.2  & 31.3  & 35.7  & 31.2  & 44.9   & 44.8\\
5522.45 & 26.0 &  4.21  &  -1.56 & 78.5  & 78.7  &  79.0  &  83.2  & 69.7  & 68.3  & 70.9  & 84.7   & 83.0
\enddata
 \tablenotetext{1}{BD+12 1917 }  
 \tablenotetext{2}{BD+12 1926} 
   \tablenotetext{3}{NGC 2423 3}
   \tablenotetext{4}{NGC 2423 15} 
   \tablenotetext{5}{NGC 2423 73} 
   \tablenotetext{6}{NGC 4349 5} 
   \tablenotetext{7}{NGC 4349 9} 
   \tablenotetext{8}{NGC 4349 127} 
   \tablenotetext{9}{NGC 4349 174} 
\tablecomments{(This table is available in its entirety in a machine-readable form in the online journal. A portion is shown here for guidance regarding its form and content.)}
\end{deluxetable*}

\begin{deluxetable*}{lcccccrrcr}
\tablecolumns{9}
\tablewidth{0pt}
\tabletypesize{\scriptsize}
\tablecaption{Measured Stellar Parameters and Abundances \label{tab:results}}
\tablehead{
   \colhead{Star} &
     \colhead{\teff} &
   \colhead{$\log g$} &
   \colhead{[Fe/H]} &
   \colhead{$\sigma_{\rm [Fe/H]}$} &
   \colhead{$\xi$} & 
   \colhead{\ali$_{\rm LTE}$} &
   \colhead{\ali$_{\rm NLTE}$} &
   \colhead{\cratio} &
   \colhead{\vsini} \\
   \colhead{  } &
      \colhead{(K)  } &
   \colhead{(dex)  } &
    \colhead{(dex)} &
    \colhead{(dex)} &
   \colhead{(\kms)} &
   \colhead{(dex)  } &
   \colhead{(dex)  } &
    \colhead{} &
   \colhead{(\kms)} }
\startdata
NGC~2423~15  &4600.   &  2.25   & $-0.04$ & 0.11 &  1.56&	$<$$-0.39$ &$<$$-0.11$   & 19 & $<2$ \\
NGC~2423~3	 &4600.   &  2.35   & $-0.02$ & 0.11 &  1.51&  +1.30    & +1.56      &  24 &	$<2$ \\
NGC~2423~73	 &5020.   &  2.95   & +0.01  &0.10 &   1.42& +0.16  &+0.32      	  &  17 &$<2$ \\
NGC~4349~127 &4370.   &  1.70   & $-0.20$&  0.11 &  1.70 &	+0.94 &+1.26       &  23 & $<2$ \\
NGC~4349~174 &4570.   &  2.05   & $-0.15$&  0.11 &  1.69 &	+0.59 &+0.87       &  21 & $<2$ \\
NGC~4349~5	 &5030.   &  2.50   & $-0.06$  & 0.10&  1.62 & $<$+0.49    &$<$+0.66   &  20 & 4.7 \\
NGC~4349~9	 &4960.   &  2.35   & $-0.18$  &0.11 &  1.53 & +1.00    & +1.18      &  20 & 7.5 \\
BD+12~1917   &4330.   &  2.06   & $-0.06$   & 0.12&  1.55 & $-0.81$    & $-0.50$   & 16 & $<2$ \\
BD+12~1926   &4300.   &  2.03   & $-0.07$   &0.12 &  1.53 & $-0.82$  &  $-0.50$ & 14 & $<2$ 
\enddata
\end{deluxetable*}

\begin{deluxetable*}{lllll}
\tablecolumns{5}
\tablewidth{0pt}
\tabletypesize{\scriptsize}
\tablecaption{Stellar Parameter Comparison \label{tab:comp}}
\tablehead{
   \colhead{\teff} &
   \colhead{$\log g$} &
   \colhead{[Fe/H]} &
   \colhead{$\xi$} &    \colhead{Source}  \\
   \colhead{(K)  } &
   \colhead{(dex)  } &
    \colhead{(dex)} &
   \colhead{(\kms)} & \colhead{}}
\startdata
\multicolumn{5}{l}{NGC  2423 3} \\
\hline \\
 $4600\pm72$  & $2.35\pm0.16$ & $-0.02\pm0.11$ & $1.51\pm0.06$ & this work \\
 $4545\pm71$  & $2.20\pm0.20$ & $-0.08\pm0.05$ & $1.55\pm0.07$ &   Mortier et~al.\ (2013)  \\ %\cite{2013A&A...557A..70M} \\
 $4680\pm100$ & $2.55\pm0.20$ & $+0.00\pm0.13$ & $1.67\pm0.20$ & \cite{2010ApJ...725..721G} \\
\hline \\
\multicolumn{5}{l}{NGC  4349 127} \\
\hline \\
 $4370\pm78$  & $1.70\pm0.22$ & $-0.20\pm0.11$ & $1.70\pm0.06$ & this work\\
 $4445\pm87$  & $1.64\pm0.23$ & $-0.25\pm0.06$ & $1.84\pm0.08$ & Mortier et~al.\ (2013) \\% \cite{2013A&A...557A..70M} \\
 $4519\pm100$ & $1.92\pm0.2$ &   $-0.21\pm0.12$ & $2.08\pm0.20$ & \cite{2010ApJ...725..721G} \\
\hline \\
\multicolumn{5}{l}{BD+12 1917} \\
\hline \\
 $4330\pm83$  & $2.06\pm0.21$ & $-0.06\pm0.13$ & $1.55\pm0.07$ & this work\\
 $4284\pm9$  & $2.2\pm0.06$ & $-0.02\pm0.04$ & \nodata &  Brucalassi et~al.\ (2014) % \cite{2014A&A...561L...9B} \\
\enddata
\end{deluxetable*}

\subsection{Lithium and \cratio}
Both \ali\ and \cratio\ were measured via spectral synthesis using the line lists and general procedure of \cite{carlberg12}. The Li line list was originally published by \cite{2009ApJ...698..451G}.  We adopted scaled-solar abundances for each star, using the solar values in  \cite{2009ARA&A..47..481A}. Since  RGs have completed first dredge-up, their C/N are not solar, but the total abundance of C+N is conserved. We adopted a ratio of C/N$=1.5$ and an initial guess of scaled-solar abundances for C+N. However, C+N was a free parameter in the synthesis fits. When fitting the \ion{Li}{1} doublet near 6708~\AA, all of our stars required an increase in the C and N abundances. We used a simple gaussian broadening to fit the spectral region, which varied from 0.15--0.26~\AA, with the exception of NGC 4349 9, which was best fit by including rotational broadening of 10~\kms\ on top of 0.15~\AA\ of gaussian broadening. The local thermodynamic equilibrium (LTE) abundances were corrected for non-LTE (NLTE) effects by interpolating the grid of corrections computed by \cite{2009A&A...503..541L}. 

 \cratio\ was measured by fitting the spectral features of $^{12}$CN and $^{13}$CN  in the 8001--8006~\AA\ spectral region.  The C+N abundances were again left as free parameters with the ratio still fixed at 1.5. Generally, the stars required 0.05~dex less of each element in the fit for \cratio\ than in the fit for \ali.

\begin{figure}[tb]
\includegraphics[width=0.5\textwidth]{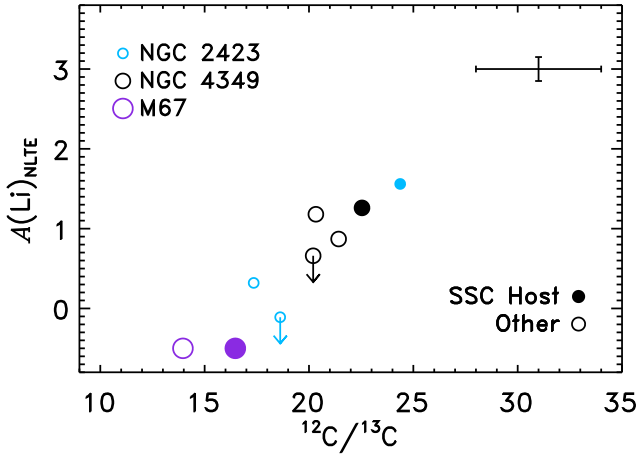}
\caption{  \ali\ versus \cratio\ for the stars in this study (circles, SSC hosts are filled symbols). The typical uncertainty is shown in the top right. \label{fig:result}}
\end{figure}

The results for \cratio\ and \ali\  are given in columns 7--9 of Table \ref{tab:results} and are plotted in Figure \ref{fig:result}.  NGC~2423~3, one of the planet hosts, is found to be Li-rich with \ali$_{\rm NLTE}$=1.56~dex.  This abundance is over an order of magnitude larger than the two comparison stars.  The brown dwarf host in NGC~4349 (star 127) also has the highest \ali\ compared to the control stars, although its abundance is consistent with that of one of the control stars (star 9) within the uncertainties.  All but one of the control stars (NGC~2423~15) in these two clusters have been monitored for substellar companions by \cite{2007A&A...472..657L}, with no companions yet found.  The two stars analyzed in M67 appear to have identical \ali, but the control star has not been monitored for substellar companions. 

There is also a tight correlation (Spearman's rank correlation coefficient, $\rho=0.96$ with $P\sim 3\times10^{-5}$) between \ali\ and \cratio\ for the stars in this study that may suggest the stars are  showing different levels of dilution. 
Excepting the Li-rich star, the correlation also progresses with cluster age (and thus stellar mass) with the youngest cluster showing the highest abundances (least dilution) and the oldest cluster showing the lowest abundances (most dilution). 
This dilution scenario is explored in more detail in Section \ref{sec:rotmixing}. 
In all cases, the SSC hosting red giant in each cluster has the largest \ali\ and \cratio\ compared to the other members we analyzed, although only NGC2423~3 has abundances that are significantly different from the other  RGs in the cluster.  
 In Section \ref{sec:evol}, we explore whether the abundance patterns are consistent with the relative evolutionary stages of the stars in each cluster. 

\subsection{\vsini}
The stellar rotational velocity (\vsini) was measured for each star by fitting profiles to six unblended iron lines as in \cite{carlberg12}. Macroturbulence  ($\zeta$) values were computed using each's stars $T_{\rm eff}$ and the \teff-$\zeta$ relationship given in \cite{2007A&A...475.1003H} for luminosity class III stars ($\sim$5~\kms).  For the NGC~2423 and M~67 stars, the combination of instrumental broadening and $\zeta$ created profiles broader than the observed lines for no rotation. For these stars, we used $\zeta$ appropriate for luminosity class IV stars ($\sim$3~\kms). 
Our measured \vsini\ are given in the last column of Table \ref{tab:results}.
We found that almost all of the stars have very slow rotation, with \vsini$<2$~\kms.  Only NGC4349~5 and NGC4349~9 have measurable rotation rates of $4.7$ and $7.5$~\kms, respectively.  Given these stars' location near the base of the RGB in the CMD in Figure \ref{fig:cmds}, this result is consistent with the picture that these stars are still spinning down.   NGC4349~9's position on the \teff-$\log g$ diagram could also be consistent with the red clump.   
Rotation models of intermediate mass stars, which leave the MS with much higher rotation rates than low mass stars, show that this rotation rate is still within the predicted red clump rotation for the fastest rotators (e.g., \citealt{2015ApJ...807...82T}).

\section{Stellar Masses}
\label{sec:masses}
One of the advantages of studying RGs in open clusters is  that their stellar ages and thus masses are better constrained than for the typical field RG.
However, changes to the inferred cluster age and metallicity affects the isochronal mass estimate for the RGs, and this is particularly problematic for the two NGC clusters whose properties are less well constrained than M67.
For the SSC hosts in these two clusters, there is also a discrepancy between the mass estimates derived purely from cluster isochrones and those that   rely on the individual stars' spectroscopic  information. Both  \cite{2010ApJ...725..721G} and   \cite{2013A&A...557A..70M}   used the online code PARAM\footnote{Available at http://stev.oapd.inaf.it/cgi-bin/param.} \citep{2006A&A...458..609D}
to estimate the likely stellar mass, age,  $\log g$, and radius of planet-hosting RGs. 
\cite{2010ApJ...725..721G}  find $2.16\pm0.38$~\msun\ for NGC~2423~3 and $3.77\pm0.36$~\msun\ for NGC~4349~127. In contrast,  \cite{2013A&A...557A..70M} find  $M=1.18\pm0.26$~\msun\ for NGC~2423~3 and $M=1.37\pm0.37$~\msun\ for NGC~4349~127. 
Some of the difference in these two calculations may result from different adoptions of reddening to the stars. 
Both studies find a lower mass for NGC2423~3 compared to the 2.4~\msun\ adopted by  \cite{2007A&A...472..657L}.  The mass of NGC4349~127 by \cite{2010ApJ...725..721G} is well within the range of previous mass estimates based on stellar isochrone fitting (3.1--3.9~\msun, see Section \ref{sec:properties}), while  \cite{2013A&A...557A..70M}  finds a substantially lower mass.

We can add an additional mass estimate  using our measured \teff\ and $\log g$ combined with literature $V$-band photometry and adopting a distance and reddening to each cluster. We adopt $V$ magnitudes from the Tycho-2 catalog \citep{2000A&A...355L..27H}, which are available for all of the NGC~2423 stars and NGC~4349~127. These are converted  to the Johnson magnitude system. Magnitudes for the remaining NGC~4349 stars come from \cite{2008A&A...485..303M}.
The adopted reddening and distances, given in Section \ref{sec:properties}, allows us to compute absolute $V$ magnitudes, $M_V$.  These are converted to bolometric luminosities ($L_{\rm bol}$) using the  bolometric corrections in Table~1 of \cite{Torres:2010gd}.
The stellar mass is then derived from $M=gL_{\rm bol}/(G4\pi \sigma_{\rm B}T_{\rm eff}^4)$, where $\sigma_{\rm B}$ is the Stefan-Boltzmann constant, and $G$ is the gravitational constant. 

We find that the NGC~2423 RGs  have masses between 1.9--2.4~\msun, with an average of 2.1~\msun. The planet hosting star has $M\sim1.9$~\msun.  The mass estimates for the NGC~4349 RGs range from 2.2--3.3~\msun, with an average of 2.8~\msun. The brown dwarf hosting star has $M\sim2.4$~\msun.
The typical total uncertainty in the mass is $\sim55\%$.
 Like the other mass estimates that rely on some spectroscopic information, our results favor lower masses than those that rely purely on photometric isochrone fitting.

\section{Discussion}
\label{sec:discuss}
\subsection{Expected \ali\ in Each Cluster}
\label{sec:evol}

In the following subsections, we discuss the relative evolutionary stages of the RGs in each cluster individually. The masses of the RGs are different enough that very different processes affecting the Li evolution are relevant for each cluster.  One crucial  difference on the RGB is whether the stars have degenerate (lower mass stars) or non-degenerate He cores (higher mass stars). 
 For solar-metallicity stars, the critical mass separating these regimes is $\sim 2.2$~\msun\ \citep{charbonnel10}.
 For stars with non-degenerate He cores, Li evolution should cease once the stars complete first dredge-up.  Stars with degenerate He cores  will evolve through both the luminosity bump and He flash stages of evolution, and both of these stages have been suggested to alter RGs' \ali\ (e.g., \citealt{1999ApJ...510..217S}, \citealt{2011ApJ...730L..12K}).  
\subsubsection{NGC 2423}
The NGC~2423 stars are near the transition mass separating degenerate and non-degenerate He cores. This fact is especially relevant for  trying to understand the enhanced \ali\ of NGC~2423~3.  Its evolutionary stage appears to be intermediate to the other two stars analyzed in the cluster and only slightly less evolved than NGC~2423~15. If  the lower mass estimates are correct, then the star may  have recently evolved through the luminosity bump, which could explain its higher \ali\  if thermohaline mixing took place at the luminosity bump (e.g., \citealt{charbonnel10}).  In this scenario, however, one would expect the similarly evolved comparison star NGC~2423~15 to also show high Li.
On the other hand, if the higher mass inferred from the cluster age is correct, then the star will transition to the core He burning stage without any thermohaline mixing taking place---as is evident by the lack of a bump feature in the isochrone in Figure \ref{fig:cmds}.  In that case, the enriched Li may be better explained as coming from an external source, such as planet engulfment.

The \ali\ results in NGC~2423 could also be described as a bifurcation or bimodal distribution of \ali, which has also been seen in NGC~752, NGC~3680, and IC~4651 (e.g., \citealt{pilachowski88}, \citealt{2001A&A...374.1017P}, \citealt{2004AJ....127.1000A}, \citealt{2004A&A...424..951P}, \citealt{2009AJ....138.1171A}).   The RGs in these clusters all have masses $\sim$1.8--1.9~\msun, within the range of mass estimates for NGC~2423 but favoring the degenerate He-core case. 
One interpretation of the bifurcation is that it is discriminating between first ascent RGs (high \ali) and red clump stars (low \ali) that are occupying the same locus on the CMDs \citep{pilachowski88}. This interpretation was used by  \cite{2004A&A...424..951P} to argue that the RGs in NGC~3680, which mostly show high \ali, are dominated by first ascent stars.  However, since red clump stars should be more common than first ascent stars,  \cite{2001A&A...374.1017P} argued that it was possible that the red clump stars in NGC~3680 were in fact the high \ali\ stars.    These studies further complicate the interpretation of NGC~2423~3 and whether its high Li is related to the presence of the planet or to its evolutionary stage.

\subsubsection{NGC 4349}
  The NGC~4349 RGs are almost certainly above the mass threshold for non-degenerate He cores. 
The most similar comparison star to the SSC host is star 174, which is less evolved than the SSC host unless 174 is on the AGB  and 127 is on the RGB.  
Star 127's \ali\ exceeds 174's even after accounting for the 0.15~dex uncertainty in the abundance, which suggests that the high Li is significant. However, if star 127 is on the AGB track and not the RGB track, it would be at the right luminosity to be undergoing an early-AGB extra-mixing phase that could replenish its Li \citep{2000A&A...359..563C}.

The evolutionary status of the other two comparison stars are more difficult to interpret.
Stars 5 and 9 have nearly identical colors and magnitude yet one has the lowest \ali\ and one the second largest \ali\ of the four stars observed. They fall on the region of the isochrone where it is most difficult to discriminate first ascent and clump stars. The higher \ali\ of star 9 may be because it is at the base of the RGB and has not yet completed first dredge-up, while star 5 is in the red clump and thus has undergone the most Li depletion.  If this interpretation is correct and the SSC host is an RGB star, than the high \ali\ of the SSC host may be significant and due to the presence of the companion.

\subsubsection{M67}
 M67's RGs are definitively in  the degenerate He core mass regime, and the two RGs studied here have evolved beyond the luminosity bump. The bump can be seen in M67's isochrone in Figure \ref{fig:cmds} at $\log g\sim2.6$ and $\log T_{\rm eff}\sim3.67$.  The Li evolution of RGs in M67 is more complex than that of the other two clusters. In addition to the possible alterations to Li at the luminosity bump, the RGs in M67 are also likely the evolved counterparts of Li-dip stars.    The Li-dip refers to a feature seen when plotting \ali\ of main sequence stars in open clusters as a function of \teff, and thus mass \citep{1986ApJ...302L..49B}.  For a narrow range of masses, \ali\ is considerable lower than stars that are slightly more or less massive. For solar metallicity, this mass is $1.38\pm0.04$~\msun\ \citep{2009AJ....138.1171A}. 
Recall that  \cite{2014A&A...561L...9B} found  $1.35\pm0.05$~\msun\ for the planet-hosting star.  This could explain why the \ali\ of the literature M67 RGs (see Section \ref{sec:RGcomp}) all have higher \ali\ than our two stars if the literature stars all evolved from stars just outside of the Li-dip.  The comparison star in this work is nearly identical to the planet host in every way. Thus, there is no evidence that the presence of the planet has affected the stellar \ali\ (though we note that the comparison star has not been monitored for companions).

\subsection{ Comparison to Other  Red Giants}
\label{sec:RGcomp}
\begin{figure}[tb]
\includegraphics[width=0.5\textwidth]{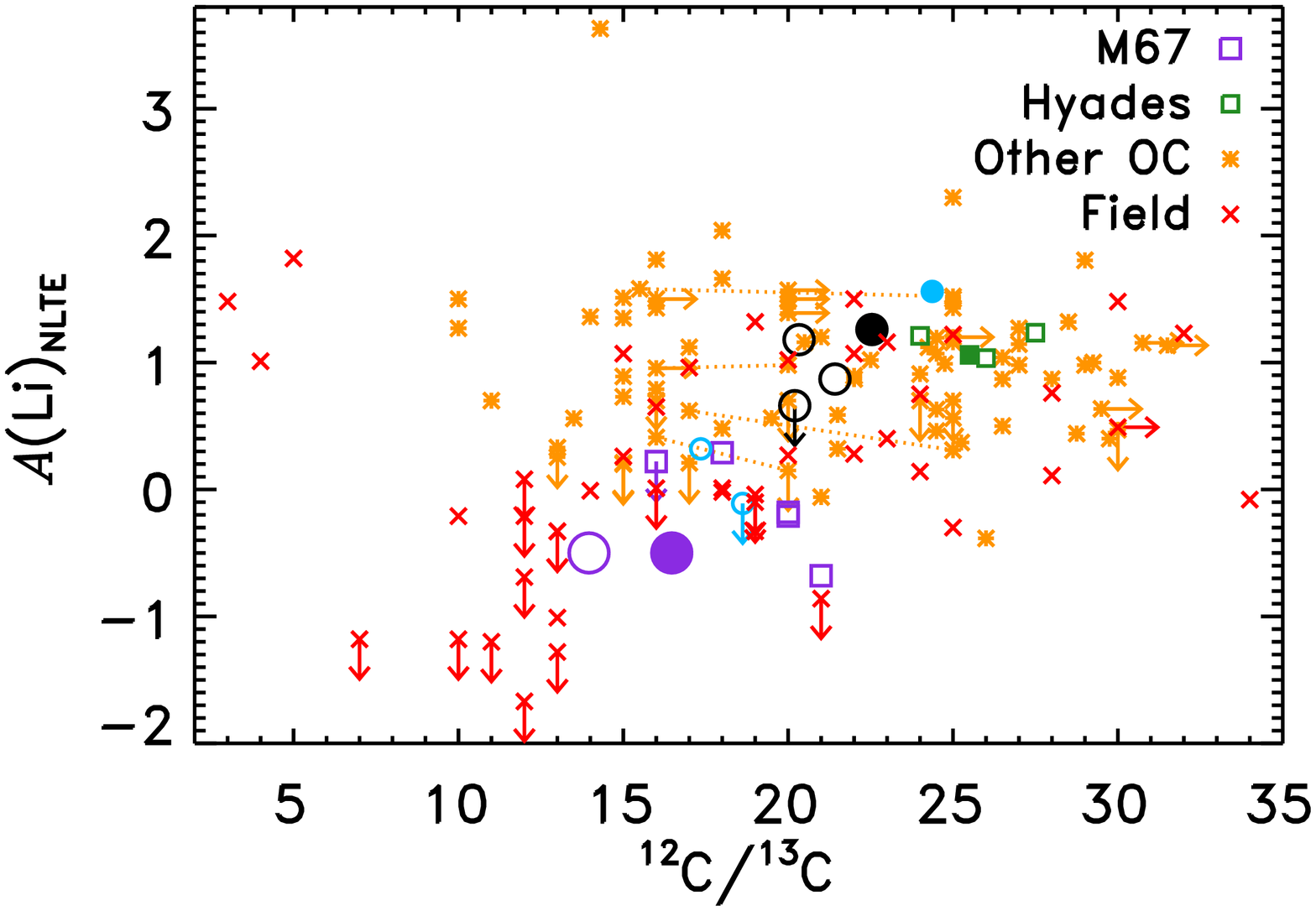}
\caption{ \ali\ versus \cratio\ for the stars in this study (as in  Figure \ref{fig:result}), with additional open cluster stars  plotted from  literature sources (\citealt{gilroy89}, \citealt{1994ApJS...91..309L}, \cite{2000PASP..112.1081G}, \citealt{2013A&A...554A...2S},  \citealt{2014A&A...564L...6M}, \citealt{2013ApJ...767L..19A}, \citealt{2015MNRAS.446.3562B}, \citealt{2015ApJ...802....7C}.).  The small, green squares highlight the Hyades, and the filled, green square is the planet host, $\epsilon$~Tau. The M67 stars from the literature (large, purple squares) combine \cratio\ from \cite{2000A&A...360..499T} with \ali\ from \cite{2011A&A...527A..94C}. 
The field giants come from \cite{1980ApJ...235..114L}.
All of the literature studies that only published LTE abundances have been corrected for NLTE effects. Dashed lines connect 4 stars in NGC~752 that were analyzed in two independent studies.  \label{fig:result2}}
\end{figure}

 In  Figure \ref{fig:result2}, we compare our measurements of \ali\ and \cratio\ to open cluster RGs and field RGs in the literature.   The field giants  come from \cite{1980ApJ...235..114L}.  Over half of the open cluster sources come from  \cite{gilroy89}, which notably includes four RGs in the Hyades  including the planet host $\epsilon$~Tau \citep{2007ApJ...661..527S}.  \cite{gilroy89} found that the mean \cratio\ of each cluster was positively correlated with the cluster turn-off mass (similar to what we find) whereas the average \ali\ showed no correlation with turn-off mass.
Although the  \cite{gilroy89} study included M67 stars, only \cratio\ was measured for them. However, there are five stars in M67 for which we have \ali\ measured by \cite{2011A&A...527A..94C}  and updated \cratio\ measured by \cite{2000A&A...360..499T}. Figure \ref{fig:result2} also includes
four NGC~6819 RGs that combine  \ali\ come from \cite{2013ApJ...767L..19A} with \cratio\ from \cite{2015ApJ...802....7C}.
The remaining stars come from  \cite{1994ApJS...91..309L}, \cite{2000PASP..112.1081G}, \cite{2013A&A...554A...2S}, \cite{2014A&A...564L...6M} and \cite{2015MNRAS.446.3562B}, where each study provided both the \ali\ and \cratio\ measurements for each star.
With the exception of the  NGC~6819 stars and the  Trumpler~5 stars \citep{2014A&A...564L...6M}, all of the literature studies presented only LTE abundances and thus have had NLTE corrections applied in the same way by us.

 Among the field stars, there is a correlation between \ali\ and \cratio\ that is especially apparent at \cratio\ $<15$.  \cite{1980ApJ...235..114L} suggested that  these  stars have masses $<1.3$~\msun.  In contrast, the open cluster stars generally have masses $>1.3$~\msun\ and  show no correlation between \ali\ and \cratio.
In the broader context of these literature abundances,  none of our stars'  \ali\ and \cratio\ appear unusual, but the tight correlation we found in Figure \ref{fig:result} is significant.  We tested the likelihood that drawing nine stars at random from either the full sample of open cluster  stars  or the large \cite{gilroy89} subsample
 would have abundances as correlated as in our sample ($\rho \geq 0.96$), and we found the probability to be low.  The likelihood is $\sim$0.2\% for the \cite{gilroy89} sub-sample and $\sim$0.03\% for the entire literature open cluster sample. 

 If the correlation between \ali\ and \cratio\ that we find is real, the lack of such a correlation in the literature open cluster
  sample may be due in part to the fact that it was not homogeneously analyzed. 
   For example, there are seven stars
  in common between \cite{2000A&A...360..499T} (adopted \cratio\ for M67 literature stars) and  \cite{gilroy89} and \cite{1991ApJ...371..578G} that we can compare. The newer measurements  find larger \cratio\  for all but one of the stars in common, with $\Delta$\cratio\ ranging from $-3$ to 14.
Alternatively, the large literature sample may simply  indicate a more complex relationship between \ali\ and \cratio\ in more massive stars. 
  Because lithium is more sensitive to the details of mixing than \cratio,  different degrees of non-standard mixing could lead to a wide spread of \ali\ at a given \cratio\  (see Section \ref{sec:rotmixing}). Additionally, any replenishment of Li due to companion engulfment will further confuse any relationship between \ali\ and \cratio\ that is due solely to stellar evolution processes.

\subsection{The Role of Non-Standard Mixing in Abundances}
\label{sec:rotmixing}
\begin{figure*}[tb]
\includegraphics[width=\textwidth]{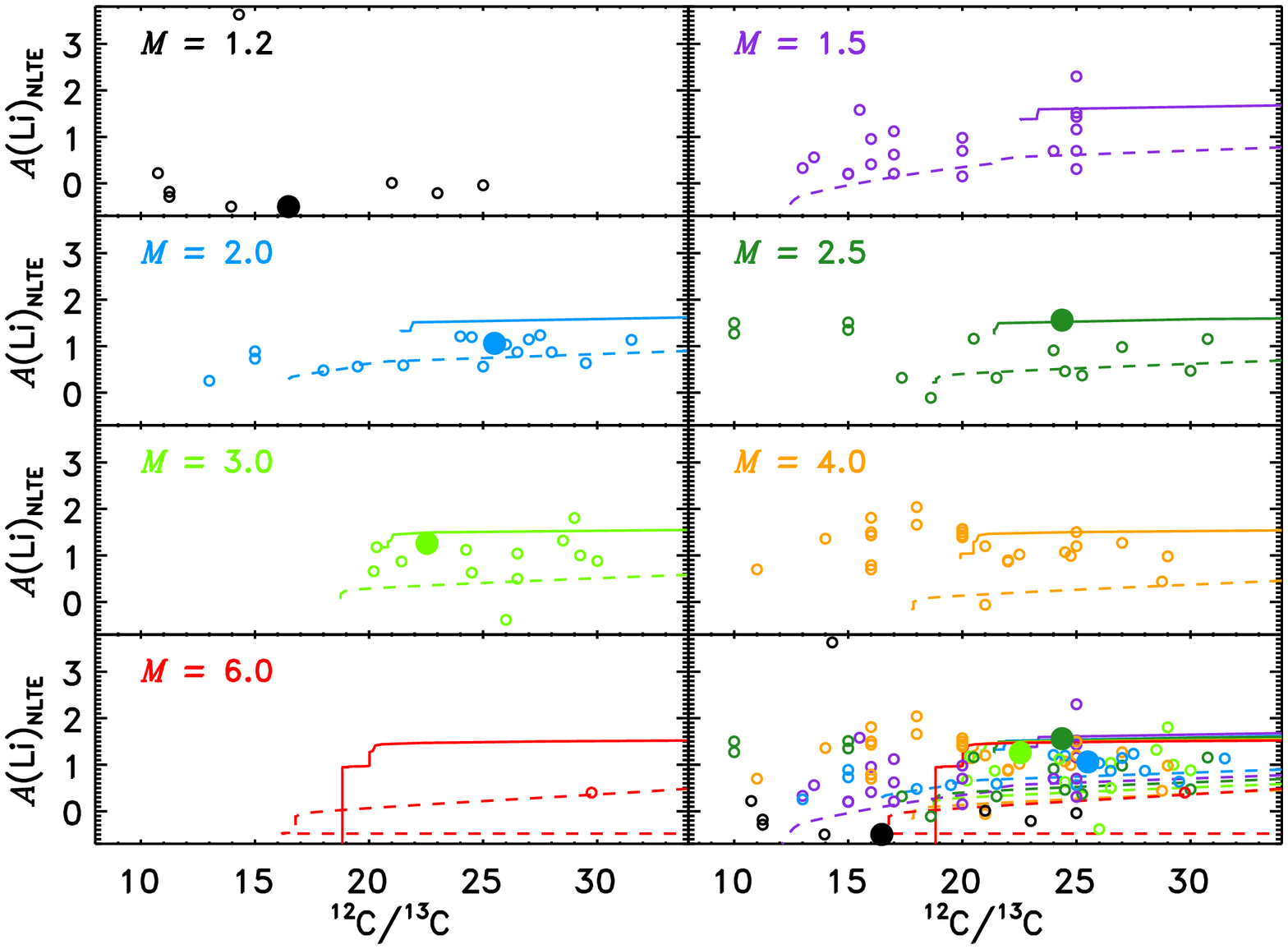}
\caption{\ali\ and \cratio\ for all of the  open cluster stars presented in Figure \ref{fig:result2} (circles, filled symbols are SSC hosts), compared to  \cite{2012A&A...543A.108L}  models 
of how these abundances change under standard mixing (solid lines) and adding thermohaline and rotation-induced mixing (dashed lines).
 The color-coding indicates the mass of the models, and the first seven panels show a single-mass model and the observed stars with turn-off masses closest to the model mass. The last panel shows all of the data and models together.
 The standard models do not include a 1.2~\msun\ model. The 1.2~\msun\ rotation model is plotted, but it predicts \ali\ a few orders of magnitude below the plot range shown.
\label{fig:models}}
\end{figure*}
The correlation between \ali\ and \cratio\ seen in Figure \ref{fig:result} suggests that the stars have experienced different levels of mixing, since both \ali\ and \cratio\ are reduced by deeper mixing.   
To explore this possibility,  we compare  the observed \ali\ and \cratio\ to model predictions from \cite{2012A&A...543A.108L} in Figure \ref{fig:models}.
 These models were run to explore the differences between models that adopt standard mixing only and those that include both thermohaline and rotation-induced mixing. The mass range of the models brackets the mass range of the stellar sample. 
The observed data are color-coded to match the closest matching model mass,  and comparisons between the individual model masses and all of the models together are shown in separate panels. The standard models reproduce well the upper envelope of \ali\ for \cratio$>20$.  
The models that include other mixing processes reproduce the lower range of \ali\ for a wider range of \cratio.
Only one initial condition for non-zero rotation was considered in these models.   \cite{charbonnel10} showed that stars with lower MS rotation rates had less rotation-induced mixing and thus higher \ali\ as RGs.  Since the majority of our stars have abundances that fall between the standard models and the rotation models, we can assume that these stars had slower MS rotation rates than  what was adopted for the models.   However, the models fail to reproduce the stars with low \cratio\ but relatively high \ali.

With the exception of the M67 stars, all of the open cluster stars with known SSCs are more massive than the Kraft break ($\sim$1.3~\msun, \citealt{1967ApJ...150..551K}) and thus have a wide range of rotation rates on the MS.  Thus, for the SSC hosts in NGC~2423, NGC~4349 and the Hyades, the relative \ali\ compared to the non-hosts may relate to different amounts of rotational mixing on the main sequence.  
 In these clusters, the SSC hosts have either considerably higher (NGC~2423) or comparable \ali\ to the other stars in the cluster; therefore, 
 we can conclude that the SSC hosts had comparable or smaller MS rotation rates than the non planet hosts. 
This interpretation is in conflict with the results of \cite{2010MNRAS.408.1770A}, who found observationally that planet hosting MS stars tend to have higher angular momentum than their planet-less counterparts, especially at masses $\gtrsim1.3$~\msun.  On the other hand, \cite{2005ApJ...631..540C}  demonstrated that the degree of rotational mixing on the RGB for the same MS turn-off rotation was sensitive to the steepness of the differential rotation profile of the star. Thus, if the planet hosts had similar turn-off rotation rates to the stars in their respectively clusters, their higher observed \ali\ and \cratio\ may also indicate a shallower rotation profile, rather than a slower MS rotation rate,
 compared to the other cluster stars.

\subsection{Li in Planet-Hosting Red Giants}
 The first Li-rich planet hosting star, BD+48 740 \citep{Adamow:2012ii}, is a field star and has \ali$_{\rm NLTE}$ of 2.33~dex (updated to 2.07~dex in \citealt{2014A&A...569A..55A}), slow rotation, but no measurement of \cratio.   The authors consider the eccentricity of the planet's orbit to be uncommonly high ($e=0.67$), and they argue that the high eccentricity and  stellar \ali\  suggested the ingestion of an inner, massive planet.
Planet ingestion could trigger a mass loss event that results in a brief period of  infrared excess \citep{1999MNRAS.308.1133S}.  BD+48 740 and the three SSC hosting stars studied here were included in a recent comprehensive  survey of IR excess around known Li enhanced stars \citep{2015AJ....150..123R}, but no excess was found for any of them. 

 We can also test whether the \ali\ of our SSC hosts correlate with eccentricity  or any other orbital property.
Table \ref{tab:PH} presents  the stellar \ali\ and companion orbital properties of all of the open cluster SSC hosts  ordered from highest to lowest stellar \ali.  These properties include the companion mass ($m\sin i$), orbital period ($P$), semi-major axis ($a$), and eccentricity ($e$).  We also list $M_\star$ and $R_\star$ and give the distance of periastron passage ($r_{\rm peri}$) as a fraction of the stellar radius. This last property gives a sense of the relative importance of possible tidal interactions between the planet and host star.  The ``Ref.'' column gives the source of the $M_\star$ and $R_\star$  measurement.   Both $m\sin i$ and $a$ have been recalculated using the adopted values of $M_\star$.
The table also includes the planet-hosting field star BD+48 740.

Although  BD+48 740 has both the largest stellar \ali\ and hosts the planet with the largest eccentricity, the trend does not continue in the open cluster stars.  The star hosting the next most eccentric planet (BD+12~1917)  has the lowest \ali\ in this sample.  The only other Li-rich star (NGC~2423~3) does not stand out from the sample in any orbital characteristic of the planet.  The \ali\ does trend with the orbital periods of the companions, with the largest \ali\ corresponding to the longest periods. However, because of the different stellar masses and orbital eccentricities, this trend does not extend to the orbital semi-major axes or the periastron passage distances of the companions. 
Based on this small sample, there appears to be no  obvious physical connection between a RG's Li and the orbital properties of the companion it hosts.

\begin{deluxetable*}{lcrrrrrrrr}
\tablecolumns{10}
\tablewidth{0pt}
\tabletypesize{\scriptsize}
\tablecaption{Orbital Parameters of Planet Hosts \label{tab:PH}}
\tablehead{
   \colhead{Name} &
   \colhead{\ali$_{\rm NLTE}$} &
   \colhead{$m\sin i$\tablenotemark{a}} &
   \colhead{$P$} &
   \colhead{$a$\tablenotemark{a}} &
   \colhead{$e$} &
   \colhead{$R_\star$} &
   \colhead{$M_\star$} &
   \colhead{$r_{\rm peri}$} &
   \colhead{Ref.}\\
   \colhead{  } &
   \colhead{ (dex)} &
   \colhead{ ($M_{\rm Jup}$) } &
   \colhead{ (days) } &
   \colhead{ (AU) } &
   \colhead{  } &
   \colhead{ ($R_\sun$) } &
   \colhead{ ($M_\sun$) } &
   \colhead{ ($R_\star$) } &
   \colhead{} }
\startdata
 & & &  & & & & & & \\
BD+48~470~b         &+2.33 & 1.6  &    $771.3\pm7.4$   & 1.88    & $0.67\pm0.17$     & $11.4\pm0.7$      & $1.5\pm0.3$      & $11.7$ & A12 \\
NGC2423~3~b       & +1.56  & $9.1$      & $714.3\pm5.3$   & 2.02      & $0.210\pm0.07$   & $14.11\pm0.88$ & $2.16\pm0.38$ & $24.3$ & G10 \\
NGC4349~127~b   &  +1.26 & $19.3$       & $677.8\pm6.2$   & 2.35       & $0.190\pm0.07$   & $44.72\pm2.46$ & $3.77\pm0.36$ & $9.2$   & G10 \\
$\epsilon$~Tau~b   & +1.06  & $7.7$      & $594\pm5.3$      & 1.94     & $0.151\pm0.023$ & $12.69\pm0.46$ & $2.75\pm0.11$ & $27.9$ & G10 \\
BD+12~1917~b      & $-0.50$  & $1.9$ & $121.71\pm0.3$ & 0.53 & $0.350\pm0.08$   & $21.8\pm0.7$     & $1.35\pm0.05$ & $3.4$   & B14 
\enddata
\tablenotetext{a}{Recalculated using adopted $M_\star$.}
\tablerefs{A12: \cite{Adamow:2012ii}, G10: \cite{2010ApJ...725..721G}, B14: \cite{2014A&A...561L...9B}}
\end{deluxetable*}

\section{Conclusions}
We have measured spectroscopic stellar parameters and abundances for nine RG stars in three open clusters,  with one substellar companion host in each cluster.  The average [Fe/H]  for the clusters are found to be
$-0.01$~dex (NGC~2423), $-0.15$~dex  (NGC~4349), and $-0.06$~dex (M67).  Our focus is on the \ali\ and \cratio\ of the RGs to explore whether the presence
of planets affects the stellar Li content either through differences in stellar mixing (traceable with \cratio) or possibly through the  past engulfment of other planets in the system.
One of the planet-hosts, NGC~2423~3, is found to be Li-rich and has significantly higher \ali\ than the two comparison stars we measured.  The brown dwarf host, NGC~4349~127, has the highest \ali\  compared to its comparison stars;  however, one of the comparison stars has a similar abundance within the uncertainties. 
 In contrast, the planet host in M67 is nearly identical to its comparison star in every stellar parameter we measured, including \ali\ and \cratio.  Similarly, the planet-hosting star in the Hyades   (a cluster for which \ali\ and \cratio\  are available in the literature) does not have any unusual abundances. 

Since Li-richness in RGs is a relatively rare phenomenon,  only one other planet-host was previously known to be Li-rich \citep{Adamow:2012ii}. Combining this star with the open cluster SSC hosts, we looked for a connection between the companion's orbital properties and the stellar \ali. We found that \ali\ correlated with the companion's orbital period, but there was no correlation to the distance of periastron passage or eccentricity  that would suggest a past planet-star interaction. Additionally, none of the stars show  evidence of an IR excess \citep{2015AJ....150..123R}.

The stellar Li content on the RGB is also largely influenced by
stellar mass and the associated differences in mixing histories.  Therefore, we have compared our \ali\ and \cratio\ to measurements of other cluster stars in the literature in an attempt to understand  the mixing history of stars and the potential relationship between \ali\ and whether the star hosts a planetary companion.
In the intermediate mass regime ($\sim$2--3.5~\msun), our planet-hosts have some of the highest \ali.  If rotation-induced mixing is the cause of the \ali\ spread for stars in this mass regime, this could imply that the planet hosts had lower MS rotation rates. 
 However, a limitation of this interpretation  is the small number of stars in each cluster that have both \ali\ and \cratio. 
Studying larger populations of stars in individual open clusters will better constrain the role of mass and evolutionary stage on \ali, providing the opportunity to elucidate the role of rotation and the presence of planets.

Another limitation of this study is the discrepant mass estimates for the cluster RGs which limits  our knowledge of the stars' current evolutionary phases and whether  they have degenerate or non-degenerate He cores.
This limitation is especially problematic for the Li-rich planet host NGC~2423~3, whose mass estimate is near the dividing mass between these two evolutionary pathways.  Its evolutionary stage would be at the luminosity bump if its core is degenerate; thus, its higher Li could also plausibly be due to thermohaline mixing.
Improved photometry for better characterization of the cluster properties and precise parallax measurements (such as what will be provided by $Gaia$  \citealt{2012Ap&SS.341...31D}) would help resolve the mass uncertainty.

\acknowledgments
This research was supported by an appointment to the NASA Postdoctoral Program at the Goddard Space Flight Center, administered by Oak Ridge Associated Universities through a contract with NASA.  This work made use of the WEBDA database, operated at the Department of Theoretical Physics and Astrophysics of the Masaryk University and the SIMBAD database, operated at CDS, Strasbourg, France.

\end{document}